\documentclass[aps, prd, superscriptaddress, nofootinbib, preprintnumbers, twocolumn, showpacs]{revtex4}

\usepackage{graphicx}
\usepackage{amsmath}
\usepackage{multirow}

\def\fsl#1{\setbox0=\hbox{$#1$}                 
   \dimen0=\wd0                                 
   \setbox1=\hbox{/} \dimen1=\wd1               
   \ifdim\dimen0>\dimen1                        
      \rlap{\hbox to \dimen0{\hfil/\hfil}}      
      #1                                        
   \else                                        
      \rlap{\hbox to \dimen1{\hfil$#1$\hfil}}   
      /                                         
   \fi}                                         %

\newcommand{\sgn}{\mbox{sgn}}

\newcommand{\VEV}[1]{\langle #1 \rangle}

\newcommand{\be}{\begin{equation}}
\newcommand{\ee}{\end{equation}}
\newcommand{\ba}{\begin{eqnarray}}
\newcommand{\ea}{\end{eqnarray}}

\begin{document}

\title{Excitonic gap, phase transition, and quantum Hall effect in graphene}

\date{\today}


\author{V.P. Gusynin}

\affiliation{Bogolyubov Institute for Theoretical Physics,
03143, Kiev, Ukraine}

\author{V.A. Miransky}

\altaffiliation[On leave from ]{
Bogolyubov Institute for Theoretical Physics,
03143, Kiev, Ukraine}

\affiliation{
Department of Applied Mathematics, University of Western
Ontario, London, Ontario N6A 5B7, Canada}

\author{S.G. Sharapov}

\affiliation{Department of Physics and Astronomy,
McMaster University, Hamilton, Ontario L8S 4M1, Canada}

\author{I.A.~Shovkovy}

\altaffiliation[On leave from ]{
Bogolyubov Institute for Theoretical Physics,
03143, Kiev, Ukraine}

\affiliation{Frankfurt Institute for Advanced Studies,
Johann Wolfgang Goethe--Universit\"at,
D-60438 Frankfurt am Main, Germany}

\affiliation{Physics Department, Western Illinois University,
Macomb, IL 61455, USA}

\date{\today}

\begin{abstract}

We suggest that physics underlying the recently observed removal of sublattice and spin
degeneracies in graphene in a strong magnetic field describes
a phase transition connected with the generation of an excitonic gap. The experimental
form of the Hall conductivity is reproduced and the main characteristics of the
dynamics are described. Predictions of the behavior of the gap as a function of temperature
and a gate voltage are made.
\end{abstract}


\maketitle

\section{Introduction}
\label{1}

The properties of graphene, a single atomic layer
of graphite \cite{Geim2004Science},
have recently attracted a lot of attention, especially after
the experimental discovery \cite{Geim2005Nature,Kim2005Nature} and
(independently of that) theoretical prediction \cite{Gusynin2005PRL,Peres2005}
of an anomalous quantization in the quantum Hall (QH) effect
(for earlier considerations
of the QH effect in graphene, see Ref. \cite{Haldane1988PRL}).
The graphene material is unique because of its band structure with
two inequivalent Dirac points at the corners of the Brillouin
zone. As a result, its low-energy excitations are described
by effective ``relativistic-like'' Dirac equation where the speed of
light is replaced by the Fermi velocity $v_F$ \cite{Semenoff1984PRL}.

These relativistic-like features of graphene are at the heart of
the anomalous integer QH effect. In this case, the filling factors are
$\nu = \pm 4(|n| + 1/2)$, where $n$ is the Landau level
index. For each QH state, a four-fold
degeneracy takes place: it is the sublattice and spin degeneracy
for the lowest Landau level (LLL) with $n = 0$ and the
valley and spin degeneracy for higher Landau levels (LLs)
with $|n| > 0$.
In the very recent experiments
\cite{Zhang2006}, it has been observed that in a strong enough
magnetic field the new QH plateaus, $\nu = 0, \pm 1$ and
$\pm 4$, occur, that was attributed to the magnetic field
induced splitting of the LLL and the $n = \pm 1$ LLs.
It is noticeable that while the degeneracy of the lowest LLL
is completely lifted, only the spin degeneracy of the $n = \pm 1$ LLs is removed.

In this paper, we suggest that the origin of the plateaus $\nu = 0$ and, especially,
$\nu = \pm 1$
is deeply connected
with a phase transition with respect to the chemical potential
$\mu$ related to the charge density of carriers
(in the experiments \cite{Geim2004Science,Geim2005Nature,Kim2005Nature,Zhang2006},
the chemical potential is tunable by a gate bias voltage $V_g$).
This phase transition is provided by dynamics responsible for creating
an excitonic gap $\Delta$ in a strong magnetic field: While at small $|\mu|$ the
excitonic gap is generated, there is no gap as $|\mu|$ becomes larger than a critical
value $|\mu_c|$ determined below.
In fact, if this scenario
is correct, the phenomenon discovered in \cite{Zhang2006}
can be interpreted as the observation of two different phases in graphene.

As will be shown below,
one of the predictions of this scenario is that
the only plateaus in the Hall conductivity $\sigma_{xy}$ are those with
$\nu = 0, \pm 1$ and $\nu = \pm 2k$, $k=1, 2,$..., i.e., the plateaus observed in
Ref. \cite{Zhang2006}. The plateau $\nu = \pm 1$ appears only if both
the spin splitting is included and the gap $\Delta$ is nonvanishing.
Another prediction is that
the excitonic gap
$\Delta$ is much smaller than the gap $\sqrt{2\hbar v_{F}^2|eB|/c}$
between the LLL and the
$n = \pm 1$ LLs. In other words, the excitonic
gap is produced by weak coupling dynamics \cite{footnote1}.
This prediction can
be checked by measuring the critical temperature at which
the $\nu = 0$ and $\nu = \pm 1$ plateaus disappear. We also succeeded in
reproducing the experimental form of $\sigma_{xy}$ obtained in \cite{Zhang2006}.

The fact that a magnetic field is a strong catalyst of
electron-hole (fermion-antifermion in field theory) pairing was
established long ago \cite{Gusynin1995PRD} (the phenomenon was
called the magnetic catalysis). The essence of this effect is
the dimensional reduction $D \to D - 2$ in the dynamics of electron-hole
pairing: In a strong magnetic  field, this pairing is mostly provided
by the LLL whose dynamics is essentially $(D - 2)$-dimensional.
The dimensional reduction leads to a strong enhancement of the
density of states. As a result, for $D = 2$, as in graphene, the pairing
dynamics in infrared becomes very strong and, for zero temperature and
zero chemical potential, an excitonic gap is generated even at the
weakest attractive interaction between electrons and holes
\cite{footnote2}.
Because of this feature, the phenomenon is robust. It was
also shown in Ref. \cite{Gusynin1995PRD}, that at large
temperature or/and charge density (chemical potential),
the excitonic gap disappears. As will be discussed below, it also
disappears for a large impurity scattering rate.

In graphene, the phenomenon of the magnetic catalysis was considered
in Refs. \cite{Khveshchenko2001PRL,Gorbar2002PRB} in connection
with an interpretation of experiments in highly oriented pyrolytic
graphite \cite{kopel1}. In particular, in Ref. \cite{Gorbar2002PRB}, the
role of temperature and chemical potential in this dynamics was
clarified in detail. Also in that paper, expressions both for
the diagonal conductivity $\sigma_{xx}$ and the Hall conductivity
$\sigma_{xy}$
at nonzero gap $\Delta$ were derived and investigated.
For further studies of this phenomenon in
graphene, see Refs. \cite{Gorbar2003PLA,Khveshchenko2004nb}.

Fig. 1 illustrates the main results of our analysis.
It shows the
spectrum and the Hall conductivity $\sigma_{xy}$ in the $n=0$ and $n=1$
LLs for four different
cases corresponding to zero (nonzero) gap $\Delta$ and spin splitting.
As one can see in Fig. 1d, when both $\Delta$ and spin splitting being
nonzero, the plateaus in $\sigma_{xy}$ observed in Ref. \cite{Zhang2006}
are reproduced. Note that the degeneracies of the LLL and higher LLs
shown in Fig. 1d are different. The physics underlying Fig. 1 will be discussed
in detail in Secs. \ref{2} and \ref{3} below.

\begin{figure}
\begin{center}
\includegraphics[width=.48\textwidth]{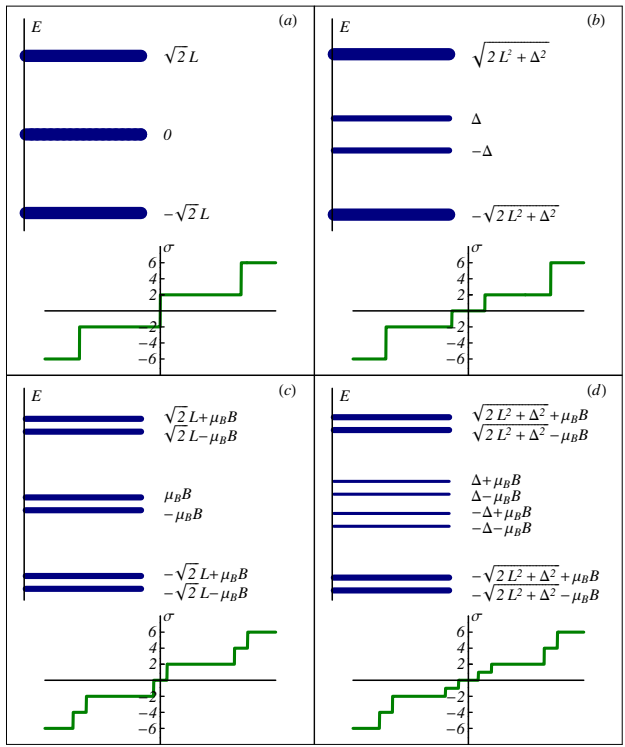}
\caption{Schematic illustration of the
spectrum and the Hall conductivity in the $n=0$ and $n=1$ Landau
levels for four different cases.
(a) $\Delta = 0$ and no Zeeman term. (b) Nonzero $\Delta$ and no Zeeman term.
(c) $\Delta = 0$ and the Zeeman term is taken into account.
(d) Both $\Delta$ and the Zeeman term are nonzero. Thickness of the
lines represents the degeneracy $\times 4, \times 2$, and $\times 1$
of the energy states; $L = \sqrt{\hbar v_{F}^2|eB|/c}$.}
\label{illustration}
\end{center}
\end{figure}

The paper is organized as follows. In Sec. \ref{2}, we consider the
dynamics of the Hall conductivity in graphene when the Zeeman term
is ignored (no spin splitting). In Sec. \ref{3}, the realistic case,
with the spin splitting taken into account, is considered. In Sec. \ref{4},
the main results of the paper are summarized. In Appendices A, B, and
C, some useful formulas and relations are derived.

\section{Dynamics with no spin splitting}
\label{2}

For the description of the dynamics in graphene, we will use the
same model as in Refs.  \cite{Khveshchenko2001PRL,Gorbar2002PRB},
the so called reduced QED. In such a model, while quasiparticles are
confined to a 2-dimensional plane, the electromagnetic (Coulomb)
interaction between them is three-dimensional in nature. The
low-energy quasiparticles excitations in graphene are conveniently
described in terms of a four-component Dirac spinor $\Psi_\sigma^T=
\left( \psi_{KA\sigma},\psi_{KB\sigma},\psi_{K^\prime B\sigma},
\psi_{K^\prime A\sigma}\right)$ which combines the Bloch states with
spin $\sigma=\pm1$ on the two different sublattices ($A,B$) of the
hexagonal graphene lattice and with momenta near the two
inequivalent points ($K,K^\prime$) at the opposite corners of the
two-dimensional Brillouin zone. The free quasiparticle Hamiltonian
can be recast in the "relativistic" - like form with the Fermi
velocity $v_F\approx 10^6 \mbox{m/s}$ playing the role of the speed
of light: \be H_0=-iv_F\int d^2{\bf
r}\,\overline{\Psi}_\sigma\left(\gamma^1\hbar\nabla_x+
\gamma^2\hbar\nabla_y\right)\Psi_\sigma, \label{free-hamiltonian}
\ee where $\overline{\Psi}_\sigma=\Psi^\dagger_\sigma\gamma^0$ is
the Dirac conjugated spinor and summation over spin $\sigma$ is
understood. In Eq.(\ref{free-hamiltonian}) $\gamma^\nu$ with
$\nu=0,1,2$ are $4\times4$ gamma matrices belonging to a reducible
representation $\gamma^\nu=\tilde{\tau}_3\otimes
(\tau_3,i\tau_2,-i\tau_1)$ where the Pauli matrices
$\tilde{\tau},\tau$ act in the subspaces of the valley
($K,K^\prime$) and sublattices ($A,B$) indices, respectively. The
matrices satisfy the usual anticommutation relations
$\left\{\gamma^\mu,\gamma^\nu\right\}=2g^{\mu\nu}$,
$g^{\mu\nu}=(1,-1,-1), \mu,\nu=0,1,2$. The covariant derivative
${\pmb \nabla}={\pmb \partial}+(ie/\hbar c){\bf A}$ includes the
vector potential in the symmetric gauge ${\bf A}^{ext}=\left(-By/2,
Bx/2\right)$ corresponding to the external magnetic field applied
perpendicular to the plane along the positive $z$ axis. In the
four-component spinor representation, the Coulomb interaction has
the form \ba H_{int}=&\frac{\hbar v_F}{2}&\int d^2{\bf r}d^2{\bf
r}^\prime\overline{\Psi}_\sigma({\bf r})\gamma^0
\Psi_\sigma({\bf r})\frac{g}{|{\bf r}-{\bf r}^\prime|}\nonumber\\
&\times&\overline{\Psi}_{\sigma\prime}({\bf r})\gamma^0
\Psi_{\sigma^\prime}({\bf r}),
\ea
where the coupling $g = {e^2}/{\epsilon_0\hbar v_F}$ and
$\epsilon_0$ is the dielectric constant (our convention
is $e > 0$).
The total Hamiltonian $H_{tot}= H_0+H_{int}$
possesses $U(4)$ symmetry discussed in Appendix \ref{C}. The chemical potential
is introduced through adding the term
$-\mu \overline{\Psi}\gamma^0\Psi = -\mu \Psi^{\dagger}\Psi$ in $H_{tot}$ (this term
preserves the $U(4)$ symmetry).
The Zeeman interaction term
is included by adding the term
$\mu_BB\overline{\Psi}\gamma^0\sigma_3\Psi = \Psi^{\dagger} \sigma_3 \Psi$,
where
now $\sigma_3$ matrix acts on spin indices. Here $\mu_B=e\hbar/(2mc)$ is the Bohr magneton and
we took into account that the Lande factor for graphene $g_L\simeq2$.

 Let us first consider a simpler case with no spin splitting
(the Zeeman term is ignored). Then, in Appendix \ref{A}, utilizing
the approach developed in Ref. \cite{Miransky:1992bj} (and used in
\cite{Gorbar2002PRB,Gorbar2003PLA}), we derive the thermodynamic
potential per unit area in a strong magnetic field $B$, when the
dynamics of the LLL dominate (here the symbol ``tilde'' in the
potential and other quantities implies that the spin splitting is
ignored):
\ba
&&\hspace{-4mm}\tilde{\Omega}(\Delta,\mu)=\frac{1}{\pi
l^2}\left\{\Delta f(\Delta,\mu)-\frac{b L(B)}{2}
f^2(\Delta,\mu)+{2T}\times\right.\nonumber\\
&&\hspace{-4mm}\left.{\rm Re}\hspace{-0.5mm}\left[\ln\Gamma\left(\frac{\gamma+i(\mu+\Delta)}
{2\pi T}+\frac{1}{2}\right)+\right.\right.\nonumber\\
&&\hspace{-4mm}\left.\left.\ln\Gamma\left(\frac{\gamma+i(\mu-\Delta)}{2\pi
T}+\frac{1}{2}\right) -2\ln\Gamma\left(\frac{\gamma} {2\pi
T}+\frac{1}{2}\right)\right]\right\}, \label{effective_pot_final}
\ea where $l =\sqrt{\hbar c/|eB|}$ is the magnetic length, $L(B) =
\sqrt{\hbar v_{F}^2|eB|/c}$ is the Landau scale, $\Gamma(x)$ is the
Euler gamma function, $\gamma$ is a LLL impurity scattering rate,
and the function $f(\Delta,\mu)$ is \be
f(\Delta,\mu)=\frac{1}{\pi}{\rm
Im}\left[\Psi\left(\frac{\gamma+i(\mu+\Delta)} {2\pi T}+
\frac{1}{2}\right) -(\Delta \to -\Delta)\right] \label{function-f}
\ee with the digamma function $\Psi(x) = \frac{d}{dx}\ln \Gamma(x)$.
The dimensionless parameter $b$ in Eq. (\ref{effective_pot_final})
reads \be \quad b=\frac{g}{\sqrt{2}}\int\limits_0^\infty\frac{dk\,
e^{-k^2}}{1+k\chi_0}, \ee where $\chi_0\simeq 0.56\sqrt{2}\pi g$.
The gap equation $\partial\tilde{\Omega/}\partial\Delta =0$ for
$\Delta$ takes the form \be \Delta-b L(B)f(\Delta,\mu) = 0.
\label{gap:eq} \ee As it is easy to see, for $T=\mu=\gamma=0$ this
gap equation has only a nontrivial solution, $\Delta=bL(B)$ (the
magnetic catalysis). For finite values of these parameters, there
exists also a trivial solution $\Delta =0$, and critical values
$T_c,\mu_c,\gamma_c$ separate the phases with zero and nonzero gaps.
The character of the phase transition can be determined by studying
the thermodynamical potential (\ref{effective_pot_final}) as a
function of these parameters. Motivated by experimental data we are
mostly interested in the phase transition with respect to the
chemical potential $\mu$, because it is easily tuned by a gate
voltage. The numerical study shows that this phase transition can be
either a first order or a second order one, depending on the values
of the scattering rate $\gamma$ and $B$. For large enough $B$ (or
small enough $\gamma$), it becomes a strong first order phase
transition.

Let us show how the generation of the gap
affects the form of the Hall conductivity $\tilde{\sigma}_{xy}$. Its expression at nonzero $\Delta$
was derived in Ref. \cite{Gorbar2002PRB}. Here we will use a compact expression
for $\tilde{\sigma}_{xy}$, valid for large $B$ and convenient for numerical calculations,
obtained recently in Ref.\cite{Gusynin2006}:
\begin{eqnarray}
\label{B.Hallcond-Delta}
&&\tilde{\sigma}_{xy} =- \frac{2e^2{\rm sgn}(eB)}{\pi h}{\rm
Im}\left\{ \Psi\left(\frac{\gamma_{\rm tr}+i(\mu+\Delta)}{2\pi
T}+\frac{1}{2}\right) \right.\nonumber\\
&&- \left.\frac{\gamma_{\rm tr}}{2\pi T}
\Psi^\prime\left(\frac{\gamma_{\rm tr} +i(\mu+\Delta)}{2\pi
T}+\frac{1}{2}\right)+ (\Delta \to - \Delta) \right\},
\label{dcHall}
\end{eqnarray}
where $\gamma_{\rm tr}$ is the transport scattering rate (for convenience of
the readers, the derivation of this expression is presented in
Appendix~\ref{B}). Note that in graphene, due, for example, to a
suppression of the backward scattering, $\gamma_{\rm tr}$ can be
smaller than the scattering rate $\gamma$ in the thermodynamic potential
(\ref{effective_pot_final}) \cite{Ando2005JPSJ}. The value of
$\gamma_{\rm tr}$ controls the sharpness of the transitions between
plateaus in the Hall conductivity.

To illustrate the role of opening the gap, let us consider the clean
limit case $\gamma= \gamma_{\rm tr} =0$ at zero temperature, when
Eq. (\ref{dcHall}) takes the form
\be \label{Hall-clean}
\tilde{\sigma}_{xy}=-\frac{2e^2}{h}{\rm sgn}\mu\,{\rm
sgn}(eB)\theta\left[|\mu|-\Delta(\mu,B)\right]. \ee When there is no
gap, this expression yields the first plateau $\nu=\pm2$ in the
half-integer QH effect
\cite{Geim2005Nature,Kim2005Nature,Gusynin2005PRL,Peres2005}. The
appearance of the gap changes the situation: In this case, for
$|\mu|<\Delta$, the
additional plateau $\tilde{\sigma}_{xy}=0$ occurs, in
accordance with the experimental data in Ref. \cite{Zhang2006}.
Note that the presence of the gap $\Delta$ leads to splitting only
the LLL. The degeneracy of higher LLs remains the same:
For these levels, the gap changes only the dispersion
relations, $E = \pm \sqrt{2\hbar v_{F}^2|neB|/c + \Delta^2}$ (compare Fig. 1a
and Fig. 1b).
Therefore, besides the plateau $\nu = 0$, other plateaus are the same
as those in the case with $\Delta = 0$, i.e., the filling factors for
them are $\nu = \pm4(|n| + 1/2)$ (see Fig. 1b).
In fact, as will be
shown in the next section, in this dynamics, the gap
disappears for the values of the chemical potential corresponding
to higher LLs, $|n| \geq 1$. Then the fact that $\nu = \pm4(|n| + 1/2)$
for these levels becomes even more evident.

Note that
lowering the degeneracy of the LLL with
generating the gap in graphene is connected with
the spontaneous breakdown of the initial $U(4)$ symmetry down to
$U(2)_{a}\times U(2)_{b}$, which is described in Appendix C.
The point is that the generation of the gap $\Delta$ is connected
with the order parameter
$\sigma = -\VEV{\bar{\Psi}\Psi} \equiv -\langle0|\bar\Psi\Psi|0\rangle$
\cite{Gusynin1995PRD,Khveshchenko2001PRL,Gorbar2002PRB} and, as shown in Appendix C,
it is invariant only under the
$U(2)_{a}\times U(2)_{b}$ subgroup of the $U(4)$.

The following remark is in order. Expression
(\ref{B.Hallcond-Delta}) does not lead to the precise values of the
Hall conductivity on the plateaus for any finite value of
$\gamma_{\rm tr}$. This happens because localized states are
neglected in our consideration. With these states taken into
account, $\gamma_{\rm tr}$ should become equal zero between Landau
levels (in particular, because of that, Eq.~(\ref{Hall-clean})
yields the correct values of the quantized Hall conductivity). Thus,
strictly speaking, with expression (\ref{B.Hallcond-Delta}), one can
describe the Hall conductivity $\sigma_{xy}$ only between plateaus,
where $\gamma_{\rm tr}$ controls the sharpness of the transitions
between the neighbor plateaus. Note however that if $\gamma_{\rm
tr}$ is small in comparison with typical energy scales in the
problem, the plateaus described by Eq. (\ref{B.Hallcond-Delta}) are
rather flat and sharp (see Fig. \ref{sigmaHall-allB} in Sec. \ref{3}
below). As is shown in Sec. \ref{3}, taking $\gamma_{\rm tr} \approx
\gamma/3$, the behavior of Hall conductivity described by Eq.
(\ref{B.Hallcond-Delta}) resembles experimental results in Ref.
\cite{Zhang2006}.

Another way to estimate possible values of $\gamma_{\rm tr}$ and
$\Delta$ is to use the expression
\begin{equation}
\label{sigma_xx-gap} \tilde{\sigma}_{xx} = \frac{2e^2}{\hbar
\pi^2} \frac{\gamma_{\rm tr}^2}{\gamma_{\rm tr}^2 + \Delta^2}
\end{equation}
for the diagonal conductivity
derived originally in Ref.~\cite{Gorbar2002PRB} and valid at the
Dirac point, $\mu=0$. Although for $\Delta=0$ it yields
$\tilde{\sigma}_{xx}= 2e^2/(\hbar \pi^2)$, which is $\pi$ times
less than the experimentally observed value in Ref. \cite{Geim2005Nature},
in the presence of $\Delta$, this expression for
$\tilde{\sigma}_{xx}$ shows a tendency towards an insulating
behavior, which is
in accordance with the data in Ref. \cite{Zhang2006}.

\section{Dynamics with spin splitting}
\label{3}

\begin{figure}
\begin{center}
\includegraphics[width=.48\textwidth]{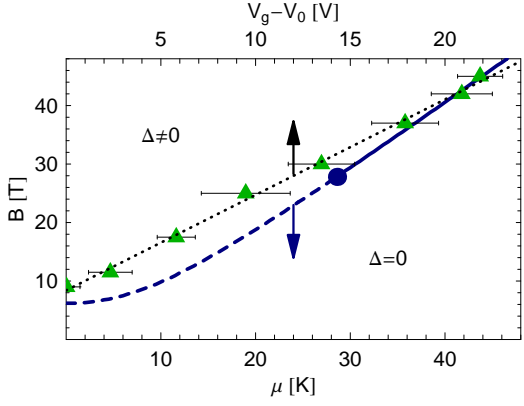}
\caption{The upper line with triangles: The experimentally observed transition points
between the lowest plateaus in the Hall conductivity from Ref.~\cite{Zhang2006}.
The lower line: The theoretical phase diagram in the plane of the
magnetic field $B$ and the chemical potential $\mu$ for $\gamma = 18$ K
and $T= 30$ mK.
First and second order phase transitions are denoted by solid
and dashed lines, respectively.}
\label{phase-diag}
\end{center}
\end{figure}

\begin{figure}
\begin{center}
\includegraphics[width=.48\textwidth]{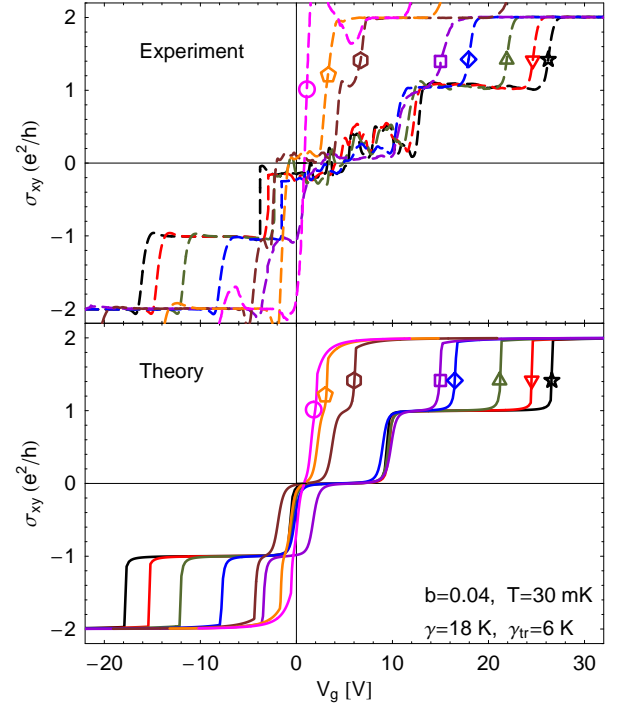}
\caption{Hall conductivity in the experiment \cite{Zhang2006}
(upper panel) and in
the theoretical model (lower panel) for magnetic fields
$B=9$~T (circle), $11.5$~T (pentagon), $17.5$~T (hexagon),
$25$~T (square), $30$~T (diamond),  $37$~T (up triangle), $42$~T (down triangle), and $45$~T (star).
The parameters in the model are $b=0.04$,
$\gamma=18$~K, $\gamma_{\rm tr}=6$~K, and temperature $T =30$ mK.}
\label{sigmaHall-allB}
\end{center}
\end{figure}

\begin{figure}
\begin{center}
\includegraphics[width=.4\textwidth]{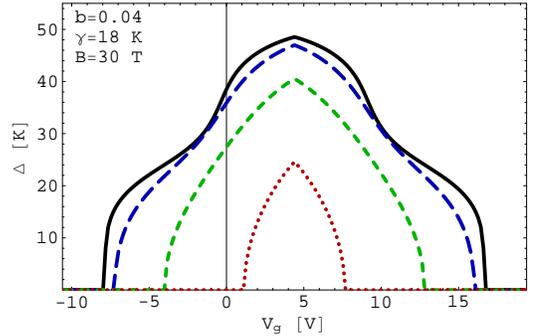}
\caption{The gap $\Delta$ versus $V_g$ for $B = 30$ T and four different values of
temperature, $T = 30$ mK, $5$ K, $10$ K, and $15$ K, from top to bottom.}
\label{Tgap}
\end{center}
\end{figure}

In the rest of the paper, we will analyze the realistic case, with a
spin splitting taken into account, whose dynamics is much richer.
The thermodynamic potential $\Omega$, the Hall conductivity
$\sigma_{xy}$ and the diagonal conductivity $\sigma_{xx}$ are now:
\ba \Omega(\Delta,\mu) = \frac{1}{2}(\tilde{\Omega}(\Delta,\mu_+) +
\tilde{\Omega}(\Delta,\mu_-)),
\label{effective_pot_final1}\\
\sigma_{xy} = \frac{1}{2}(\tilde{\sigma}_{xy}(\mu_+) + \tilde{\sigma}_{xy}(\mu_-)),
\label{dcHall1}\\
\sigma_{xx} = \frac{1}{2}(\tilde{\sigma}_{xx}(\mu_+) + \tilde{\sigma}_{xx}(\mu_-)),
\label{sigma_xx-gap1}
\ea
where the expressions for $\tilde{\Omega}$, $\tilde{\sigma}_{xy}$ and $\tilde{\sigma}_{xx}$
are given in
Eqs. (\ref{effective_pot_final}),
(\ref{dcHall}), and (\ref{dc-sigma_xx-largeB}) in Appendix \ref{B}, respectively, and
$\mu_\pm = \mu \pm \delta_z$, with $\delta_z = \mu_B B$ being the Zeeman term.
This term breaks
explicitly the $U(4)$ symmetry down to $U(2)_c \times U(2)_d$ discussed in
Appendix C.

In order to determine the parameters of the theoretical model,
we use the experimental data of Ref.~\cite{Zhang2006} as a
guide. In the regime when the excitonic gap is sufficiently
large to remove the degeneracy of the LLL, the model predicts
that one of the transition points between the plateaus in the
Hall conductivity corresponds to a phase transition in which $\Delta \to 0$ (see the discussion
connected with Fig. \ref{Tgap} below).
More precisely, it is the transition point between the $\nu=0$ and $\nu=2$ plateaus
for $B = 11.5$ T, and the transition point between the $\nu=1$ and $\nu=2$ plateaus
for larger values of $B$ (for $B = 9$ T, with no plateau $\nu = 0$,
the transition point is taken to be zero). This
is controlled by the value of the chemical potential $\mu$ (see Fig. \ref{phase-diag}).
In the experiment, the corresponding control parameter is the
gate voltage $V_g$. To obtain a relation between
the two, we compare the theoretical phase diagram on the
$B$--$\mu$ plane
with the experimental phase diagram
on the $B$--$V_g$ plane in Fig.~\ref{phase-diag}.
The latter
is obtained by a simple
compilation of the
experimentally
observed transition points between the corresponding plateaus in
the Hall conductivity.
The best fit that was found is:
\be
\mu =0.5 (V_g-V_0) + 7.0\,  \sgn(V_g-V_0) \sqrt{|V_g-V_0|},
\label{fit}
\ee
where while $\mu$ is measured in kelvins, $V_g$ is measured in
volts, and
$V_0$ is the center point in the dependence of the
Hall conductivity obtained in experiment. Note that the values of
$V_0$ are different for different values of $B$ and
are in the interval from $0.8 - 5.8$ V \cite{footnote}.

In Fig. \ref{sigmaHall-allB}, we present the experimental data for
$\sigma_{xy}$ and their description in this model for the values
of the parameters indicated in the figure. The form of the
experimental and theoretical curves are quite similar. Note
that the value of the parameter $b = 0.04$ corresponds to
a weak coupling with $g =e^2/\epsilon_0\hbar v_F \simeq 0.07$.
Because of the three dimensional nature of the Coulomb interaction,
it is plausible that the value of $g$ is influenced by a large
dielectric constant $\epsilon_0$ of the substrate in the
experimental device. The main reason of the necessity of a weak
coupling for the fit is that the lengths of the experimental plateaus
with $\nu = 0, \pm 1, \pm 2$ imply that the Zeeman energy and the
gap $\Delta$ are of the same order, and the Zeeman energy is only
$\sim$ $10$ K at $B \sim 10$ T.
Note that although there is no
$\nu =0$ plateau in the $B = 9$ T curve, the gap $\Delta$ in
this case is also nonzero (although small).
The reason is that in the presence of a nonzero scattering rate $\gamma$,
the effect of a small gap, $\Delta\alt\gamma$, is
unobservable.

In Fig. \ref{Tgap}, the gap $\Delta$ versus $V_g$ for $B = 30$ T and four different values of
temperature is shown. At low $T = 30$ mK, a strong first order phase transition with
respect to $V_g$, i.e. $\mu$, is clearly seen. It corresponds to the transition point between the
$\nu = \pm 1$ and $\nu = \pm 2$ plateaus. The steep descents, which occur approximately at
the half of the critical value of $V_g - V_0$, are related
to the transition between the $\nu = 0$ and $\nu = \pm 1$ plateaus. The phase
transition with respect to temperature is a second order one with
the critical temperature $T_c \simeq 17.2$ K.

Fig. 1 in Introduction summarizes the main results of our analysis.
Fig. 1d clearly shows that an excitonic gap $\Delta$ and a large
enough Zeeman term ($\delta_z \gtrsim \gamma$) together lead to the
QH plateaus with the filling factors $\nu = 0, \pm 1$ and $\nu = \pm
2k$, $k=1, 2,$..., i.e., those observed in Ref. \cite{Zhang2006}.
Let us discuss this point in more detail. It is noticeable that
while a large enough Zeeman term leads to the plateau $\nu = 0$ even
for $\Delta = 0$ (see Figs. 1c), the $\nu = \pm 1$ plateaus appear
only if both the Zeeman term is included and the gap $\Delta$ is
nonvanishing (compare Fig. 1d with Figs. 1b and 1c). Therefore the
$\nu = \pm 1$ plateaus are the clearest signature of the presence of
a dynamical excitonic gap. There are of course also the plateaus
$\nu = \pm 2$ connected with the LLL. As was already pointed out in
Sec. \ref{2}, the degeneracy of higher LLs is removed only by the
Zeeman term (and not by the gap $\Delta$). Therefore the filling
factors of the plateaus with $|\nu| > 1$ are described by $\nu = \pm
2k$, $k=1, 2,$....(see Fig. 1d) \cite{footnote3}. Thus, in this
scenario, the LLL plays a very special role: While the excitonic gap
does not reduce the degeneracy of higher LLs, it leads to splitting
the LLL. This point is at heart of reproducing the QH effect data
\cite{Zhang2006} in this scenario.

This picture
is intimately connected with the removal of the degeneracy in this dynamics.
As is shown in Appendix C, a nonzero $\Delta$ and
the Zeeman term together break the initial non-abelian $U(4)$ symmetry down to
the abelian $U(1)_1\times U(1)_2 \times  U(1)_3 \times U(1)_4$
one. Since irreducible
representations of abelian symmetries are one-dimensional, the $U(4)$ degeneracy
of the LLL is completely removed.

By using Eqs. (\ref{sigma_xx-gap1}) and (\ref{dc-sigma_xx-largeB}),
we also checked that the values of $\Delta$ and $\gamma_{\rm tr}$
utilized in this section yield the diagonal conductivity
$\sigma_{xx}$ whose behavior is in a qualitative agreement with the
data in Ref. \cite{Zhang2006}.

\section{Conclusion}
\label{4}

We believe that
the observation of the $\nu = 0$ and $\nu = \pm 1$ plateaus in the experiment
\cite{Zhang2006} strongly suggests that the existence
of a dynamical excitonic gap (or gaps) in graphene in a strong magnetic
field is a viable possibility.
In this paper, only a singlet excitonic gap was considered. There of course
exist other options: For example, the
same Coulomb interaction can lead also to a nonzero triplet order parameter
$\VEV{\bar{\Psi}\sigma^{3}\Psi}$. In that case, the states with up and down spins
will have different gaps, $\Delta_+$ and $\Delta_-$, respectively.
The same arguments
as those used above show that the degeneracy of higher LLs is lowered
only by the Zeeman term and, therefore, in that case the filling factors
of the the QH plateaus will be the same as in the case of the singlet excitonic gap.
Therefore the modification of Fig. 1d will be only in replacing $\Delta$
with $\Delta_{\pm}$ for up and down spins.

In the present scenario, the weak coupling dynamics was utilized.
The reason of its necessity
is that the lengths of the experimental plateaus
with $\nu = 0, \pm 1, \pm 2$ imply that the Zeeman energy and the
gap $\Delta$ are of the same order, and the Zeeman energy is only
$\sim$ $10$ K at $B \sim 10$ T. Note, however, that this argument is
valid only for a paramagnetic regime in which there is no large
enhancement of spin splitting by dynamics in a magnetic field.
When such a enhancement takes place \cite{Ferro1},
strong coupling dynamics might lead to a good fit of the
experimental data in QH effect in graphene.
This possibility will be considered elsewhere.

It is instructive to compare the present approach with
other ones used for the description of the dynamics in QH effect in
graphene.
In Refs. \cite{Ferro2,{Ferro3}}, the QH
ferromagnetism was considered. The main prediction in
Ref. \cite{Ferro2} is that in this case the QH plateaus with
{\it all} integer values of the filling factor $\nu$ occur [the
critical values of a magnetic field $B$ at which plateaus
occur are different for different $\nu$ and increase with $\nu$].
This prediction is quite different from the
present one that reflects a difference between the dynamics in these
two scenarios. As was emphasized above, the excitonic gap does not
reduce the degeneracy of higher LLs and it is
unlike the QH ferromagnetism. As a result, there are no odd filling
factors $\nu = 2k +1$, $k \geq 1$, in the scenario with excitonic
gaps.

In Ref. \cite{Para}, a scenario with the paramagnetic regime was considered
(with no large enhancement of spin splitting).
Unlike the present scenario, the breakdown of the $U(4)$ in
\cite{Para} is not
spontaneous but explicit, provided by local (on-site) interactions. The main
conclusion of Ref. \cite{Para} is that Zeeman splitting together with
on-site interactions can produce QH states at $\nu = 0, \pm 1$ and $\pm 4$
but not at $\nu = \pm 3$ and $\pm 5$. Although the values of filling factors
agree with ours,
these two dynamics are very different and should lead to very different
spectra of collective excitations.

Which of these scenarios is realized in graphene is an open issue.
It would be interesting to include all possible competing orders in the
thermodynamic potential and find the genuine ground state and the phase
diagram in graphene \cite{Bascones2002PRL}.

\acknowledgments

We are grateful to P. Kim and Y. Zhang for providing the data in the
experiment \cite{Zhang2006} and clarifying discussions. Useful
discussions with J.P. Carbotte, A.K. Geim, I.F. Herbut, D.V.~Khveshchenko, V.M.
Loktev, and Yu.G. Pogorelov are acknowledged. The work of V.P.G. was
supported by the SCOPES-project IB7320-110848 of the Swiss NSF. The
work of V.A.M. was supported by the Natural Sciences and Engineering
Research Council of Canada. A part of this work was done when V.A.M.
visited Nagoya University. He thanks Koichi Yamawaki for his
hospitality and the Mitsubishi Foundation for its financial support.
The work of S.G.S. was supported by the Natural Sciences and
Engineering Research Council of Canada and the Canadian Institute
for Advanced Research. The work of I.A.S. was supported in part by
the Virtual Institute of the Helmholtz Association under grant No.
VH-VI-041, by the Gesellschaft f\"{u}r Schwerionenforschung (GSI),
and by the Deutsche Forschungsgemeinschaft (DFG).

\appendix
\section{Thermodynamic potential}
\label{A}

In this Appendix we derive expression (\ref{effective_pot_final})
for the thermodynamic
potential per unit area $\tilde{\Omega}$
valid in the strong
field limit, $\sqrt{\hbar v_{F}^2|eB|/c} \gg \gamma, T,\mu,\Delta$.
We will use the formalism of the effective action introduced and
developed in classical papers \cite{potential, potential1}. In
those papers, the case of the effective action for elementary fields
was considered. The case of the effective action for local composite
fields was studied in Ref. \cite{Miransky:1992bj} and in our
derivation we will follow that approach.

We start from a general definition of the effective action in a
theory in which the spontaneous symmetry breaking phenomenon is
driven by the local composite order parameter
$\sigma = -\VEV{\bar\Psi\Psi} \equiv -\langle0|\bar\Psi\Psi|0\rangle$
corresponding to the generation
of the excitonic gap $\Delta$ \cite{Khveshchenko2001PRL,Gorbar2002PRB}.
Following the conventional way
\cite{potential, potential1, Miransky:1992bj},
we introduce the generating
functional $W(J)$ for the Green functions of the corresponding
composite field through the path integral
\be
e^{iW(J)}=\hspace{-1mm}\int\hspace{-1mm} D\Psi D\overline{\Psi}
\exp\left\{i\int d^3x\left[{\cal
L}_{qp}-J(x)\overline{\Psi}(x)\Psi(x)\right]\right\},
\ee
where $J(x)$ is the source for the composite field $-\bar{\Psi}(x)\Psi(x)$
and ${\cal L}_{qp}$ is the
Lagrangian density of quasiparticles in the model at hand
(note that here $x^0$ is the time variable $t$).
Then,
by definition, the effective action for the field
$\sigma(x) = -\VEV{\overline{\Psi}(x)\Psi(x)}$ is given by the Legendre
transform of the generating functional $W(J)$, \be
\Gamma(\sigma)=W(J)-\int d^3xJ(x)\sigma(x),
\label{Legendre-transform} \ee
where the external source $J(x)$ on
the right-hand side is expressed in terms of the field $\sigma(x)$
by inverting the relation \be \frac{\delta W(J)}{\delta
J(x)}=\sigma(x). \label{inverted-relation} \ee The effective
action $\Gamma(\sigma)$ in Eq. (\ref{Legendre-transform}) provides
a natural framework for describing the low energy dynamics in the
model at hand. It is common to expand this action in powers of
space-time derivatives of the field $\sigma$: \be
\Gamma(\sigma)=\int
d^3x\left[-V(\sigma)+\frac{1}{2}Z^{\mu\nu}\partial_\mu\sigma
\partial_\nu\sigma+\cdots\right],
\label{V}
\ee
where $V(\sigma)$ is the effective
potential. The ellipsis denote higher derivative terms as well as
contributions of the Nambu-Goldstone bosons.
From Eqs. (\ref{Legendre-transform}) and
(\ref{inverted-relation}), we derive the following relation: \be
\frac{\delta \Gamma(\sigma)}{\delta\sigma(x)}=-J(x).
\label{J}
\ee In the limit
of a vanishing external source, this equation turns into an
equation of motion for the composite field $\sigma(x)$. In a
particular case of constant configurations, the equation reads
$dV/d\sigma=0$.

The thermodynamic potential $\tilde{\Omega}$ per unit area in
Eq. (\ref{effective_pot_final})
is nothing else but the
effective potential $V$ at non-zero $T$ and $\mu$. The
constant source $J$ plays the role of the {\it bare} gap (Dirac mass),
$J \equiv\Delta_0$. Then the initial relation in the derivation of
$\tilde{\Omega}$ (following
from Eqs. (\ref{V}) and (\ref{J})) is:
\begin{equation}
\frac{\partial\tilde{\Omega}}{\partial\sigma} = \Delta_0.
\label{starting_point}
\end{equation}

At zero $T, \mu$ and $\gamma$, the gap equation with nonzero $\Delta_0$ in a strong field
has the form (see Eq.(51) in Ref.~\cite{Gorbar2002PRB}))
\ba
\Delta=\Delta_0&+&ie^2\int\frac{d\omega}{2\pi}\frac{\Delta}{\omega^2-\Delta^2}\nonumber\\
&\times&\int\frac{d^2k}{(2\pi)^2}\exp\left[-\frac{\hbar
c|\mathbf{k}|^2}{2|eB|}\right] U(\mathbf{k}), \ea where \be
U(\mathbf{k})=\frac{2\pi
}{\epsilon_0}\frac{1}{|\mathbf{k}|(1+a|\mathbf{k}|)}, \quad
a=4\pi\nu_0\frac{e^2}{\epsilon_0\hbar v_F}\sqrt{\frac{\hbar
c}{|eB|}}, \ee and the const $\nu_0\approx 0.14$ (see Eqs. (46) and
(47) in \cite{Gorbar2002PRB}). At finite T, $\mu$ and $\gamma$, the
gap equation is written as the following sum over Matsubara
frequencies $\omega_n=\pi T(2n+1)$: \be
\Delta=\Delta_0+2bL(B)T\sum\limits_{n=-\infty}^\infty
\frac{\Delta}{(\omega_n\hspace{-0.5mm}+\hspace{-0.5mm}\gamma\,{\rm
sgn}(\omega_n) -i\mu)^2\hspace{-0.5mm}+\hspace{-0.5mm}\Delta^2}. \ee
The sum over Matsubara frequencies is easily performed, \ba
&&\hspace{-10mm}T\sum\limits_{n=-\infty}^\infty\frac{\Delta}{(\omega_n+\gamma\,{\rm
sgn}(\omega_n)
-i\mu)^2+\Delta^2}\nonumber\\
&&\hspace{-10mm}=\frac{1}{2\pi}{\rm Im}\left[\Psi\left(\frac{\gamma+i(\mu+\Delta)}
{2\pi T}+\frac{1}{2}\right)-\left(\Delta\rightarrow-\Delta\right)\right],
\label{Matsubara-sum}
\ea
where $\Psi$ is the digamma function.
Then the gap equation takes the form
\ba
\Delta=\Delta_0&+&bL(B)\frac{1}{\pi}{\rm Im}\left[\Psi\left(\frac{\gamma+i(\mu+\Delta)}
{2\pi T}+\frac{1}{2}\right)\right.\nonumber\\
&-&\left.\Psi\left(\frac{\gamma+i(\mu-\Delta)}
{2\pi T}+\frac{1}{2}\right)\right].
\label{gap-equation}
\ea
Let us express $\Delta_0$ through $\Delta$ from this equation and
substitute it into Eq. (\ref{starting_point}). Then, taking into account
the relation
\begin{equation}
\frac{\partial\tilde{\Omega}}{\partial\sigma} =
\frac{\partial\tilde{\Omega}}{\partial\Delta} \cdot
\frac{d\Delta}{d\sigma},
\end{equation}
we come to the final equation \be
\frac{\partial\tilde{\Omega}}{\partial\Delta}
=\frac{d\sigma}{d\Delta} \left[\Delta -b L(B)f(\Delta,\mu)\right],
\label{final} \ee where the function $f(\Delta,\mu)$ is given in Eq.
(\ref{function-f}). The condition
$\partial\tilde{\Omega}/\partial\Delta=0$ yields gap equation
(\ref{gap:eq}) in the main text.

Since the field
$\sigma=-\langle\overline{\Psi}\Psi\rangle$, we need to evaluate the
chiral (excitonic) condensate which is given by
\be
\langle\overline{\Psi}\Psi\rangle=-T\sum\limits_{n=-\infty}^\infty\int\frac{d^2k}{(2\pi)^2}
\,{\rm tr}S(\omega_n,k),
\ee
where the fermion propagator in the LLL
is
\be
S(\omega_n,k)=2P_{-}e^{-k^2/|eB|}\frac{1}{\gamma_0(\omega_n+\gamma\,{\rm
sgn}(\omega_n)-i\mu)-\Delta}
\ee
with $P_{-}=(1-i\gamma^1\gamma^2)/2$.

Calculating the trace and integrating over momenta we get \be
\langle\overline{\Psi}\Psi\rangle=-\frac{2|eB|T}{\pi}\sum\limits_{n=-\infty}^\infty
\frac{\Delta}{(\omega_n+\gamma\,{\rm
sgn}(\omega_n)-i\mu)^2\hspace{-1mm}+\hspace{-1mm}\Delta^2}. \ee The
sum over Matsubara frequencies is evaluated by means of Eq.
(\ref{Matsubara-sum}) and we obtain \be
\sigma=-\langle\overline{\Psi}\Psi\rangle=\frac{1}{\pi
l^2}f(\Delta,\mu). \ee Therefore Eq. (\ref{final}) can be rewritten
as \be \frac{\partial\tilde{\Omega}}{\partial\Delta} =\frac{1}{\pi
l^2}\frac{df(\Delta,\mu)}{d\Delta} \left[\Delta -b
L(B)f(\Delta,\mu)\right]. \ee Integrating over $\Delta$ we find \ba
&&\tilde{\Omega}(\Delta,\mu)=\frac{1}{\pi l^2}\left\{\Delta
f(\Delta,\mu)-\frac{b L(B)}{2}
f^2(\Delta,\mu)\right.\nonumber\\
&&\left.+{2T}{\rm
Re}\left[\ln\Gamma\left(\frac{\gamma+i(\mu+\Delta)}{2\pi
T}+\frac{1}{2}\right)\right.\right.
\nonumber\\
&&\left.\left.+\ln\Gamma\left(\frac{\gamma+i(\mu-\Delta)}{2\pi
T}+\frac{1}{2}\right)\right]+C(\mu)\right\}, \label{effective_pot}
\ea where the integration constant $C(\mu)$ was added on the
right-hand side. Since $C(\mu)$ determines only the overall
normalization of the potential, we can take it such that
$\tilde{\Omega}(\Delta=0,\mu=0)=0$. As a result, we arrive at Eq.
(\ref{effective_pot_final}) in the main text. From Eq.
(\ref{effective_pot}) we also find that on the solution of the gap
equation $\partial\tilde{\Omega}/\partial\Delta=0$, the carrier
density $\rho$ is given by the expression \ba
&&\frac{\partial\tilde{\Omega}}{\partial\mu}=-\rho=-\frac{1}{\pi
l^2}\frac{1}{\pi}{\rm Im}
\left[\Psi\left(\frac{\gamma+i(\mu+\Delta)}{2\pi T}+\frac{1}{2}\right)\right.\nonumber\\
&&\left.+\Psi\left(\frac{\gamma+i(\mu-\Delta)} {2\pi
T}+\frac{1}{2}\right)\right].
\ea

\section{Calculation of the conductivities}
\label{B}

In the bare bubble approximation, the expression for the diagonal
conductivity in the limit of $B \to \infty$ can be obtained from
Eqs.~(3.11), (3.12) in the second paper in
Ref.~\cite{Gusynin2005PRL} ($\hbar =1$ in Appendix~B)
\begin{equation}
\label{sigma_xx-B=infty}
\begin{split} \tilde{\sigma}_{xx}& =\frac{e^2\gamma_{\rm
tr}}{\pi^2}\int\limits_{-\infty}^\infty
\frac{d\omega}{4T\cosh^2\frac{\omega-\mu}{2T}}\\
&\left[\frac{\gamma_{\rm tr}}{\gamma_{\rm tr}^2+(\omega-\Delta)^2}+
\frac{\gamma_{\rm tr}}{\gamma_{\rm tr}^2+(\omega+\Delta)^2}\right]
\end{split}
\end{equation}
[due to reasons pointed out in Sec. \ref{2}, we use $\gamma_{\rm tr}$, and not
$\gamma$, in conductivities $\sigma_{xx}$ and $\sigma_{xy}$].
The integrals in this expression can be evaluated exactly as
follows. First we write
\begin{equation}
\begin{split}
&I = \int\limits_{-\infty}^\infty
\frac{d\omega}{\cosh^2\frac{\omega-\mu}{2T}}\frac{\gamma_{\rm tr}}{\gamma_{\rm tr}^2+(\omega-\Delta)^2}\\
& = {\rm Re}\int\limits_0^\infty dt\,e^{-\gamma_{\rm tr}
t}\int\limits_{-\infty}^\infty
\frac{d\omega\, e^{it(\omega-\Delta)}}{\cosh^2\frac{\omega-\mu}{2T}}\\
&=2T{\rm Re}\int\limits_0^\infty dt\,e^{- t[\gamma_{\rm
tr}-i(\mu-\Delta)]}\int
\limits_{-\infty}^\infty\frac{dx}{\cosh^2x}\,e^{i2Ttx} \\& =4T{\rm
Re}\int\limits_0^\infty dt\,e^{- t[\gamma_{\rm
tr}-i(\mu-\Delta)]}\int\limits_{0}^\infty\frac{dx\cos(2Ttx)}{\cosh^2x}.
\end{split}
\end{equation}
The integral over $x$ is evaluated by means of the
formula (3.982.1) from Ref.~\cite{Gradstein_Ryzhik}. Then we get
\begin{equation}
I= 4\pi T^2\int\limits_0^\infty \frac{dt t\cos[(\mu-\Delta)t]\,e^{-
t\gamma_{\rm tr}}} {\sinh(\pi Tt)}. \label{important_integral1}
\end{equation}
This integral can be evaluated by differentiating Eq.~(4.131.3) of
Ref.~\cite{Gradstein_Ryzhik} which yields
\begin{equation}
\label{important_integral}
\begin{split}
&I= 4\pi T^2\int\limits_0^\infty \frac{dt t\cos[(\mu-\Delta)t]\,e^{-
t\gamma_{\rm tr}}} {\sinh(\pi Tt)}\\
& =\frac{2}{\pi}\,{\rm Re}\,\Psi^\prime\left(\frac{\gamma_{\rm
tr}+i(\mu-\Delta)}{2\pi T} +\frac{1}{2}\right).
\end{split}
\end{equation}
Thus we obtain
\begin{equation}
\label{dc-sigma_xx-largeB}
\begin{split}
\tilde{\sigma}_{xx}=& \frac{e^2\gamma_{\rm tr}}{2\pi^3T}\,{\rm
Re}\left[\Psi^\prime \left(\frac{\gamma_{\rm tr}+i(\mu+\Delta)}{2\pi
T}+\frac{1}{2}\right)\right.
\\
&+\left. \Psi^\prime \left(\frac{\gamma_{\rm tr}+i(\mu-\Delta)}{2\pi
T} +\frac{1}{2}\right)\right].
\end{split}
\end{equation}
One can check that for $\mu=0$ and $T\to 0$
Eq.~(\ref{dc-sigma_xx-largeB}) reduces to Eq.~(\ref{sigma_xx-gap}).

Now we derive the expression for the dc Hall conductivity.  In the limit
$B \to \infty$, Eqs.~(3.14) and (3.15) in the second paper in Ref.~\cite{Gusynin2005PRL}
yield
\begin{equation}
\label{sigma_xy-B=infty}
\begin{split}
\tilde{\sigma}_{xy}& =-\frac{e^2{\rm
sgn}(eB)}{4\pi^2T}\int\limits_{-\infty}^\infty
\frac{d\omega}{\cosh^2\frac{\omega-\mu}{2T}}\\
& \times \left[\frac{\gamma_{\rm tr}(\omega-\Delta)} {\gamma_{\rm
tr}^2+(\omega-\Delta)^2}+\frac{\gamma_{\rm tr}(\omega+\Delta)}
{\gamma_{\rm tr}^2+(\omega+\Delta)^2} \right. \\
&\left. +\arctan\frac{\omega+\Delta}{\gamma_{\rm tr}}
+\arctan\frac{\omega-\Delta}{\gamma_{\rm tr}}\right].
\end{split}
\end{equation}
We first consider the terms with $\arctan$ functions taking
derivative with respect to $\Delta$:
\begin{equation}
\begin{split}
\frac{d\tilde{ \sigma}_{xy}^{(1)}}{d\Delta}&=-\frac{e^2{\rm
sgn}(eB)}{4\pi^2T}\int\limits_{-\infty}^\infty
\frac{d\omega}{\cosh^2\frac{\omega-\mu}{2T}}\\
&\times \left[\frac{\gamma_{\rm tr}}{\gamma_{\rm
tr}^2+(\omega+\Delta)^2} -\frac{\gamma_{\rm tr}}{\gamma_{\rm
tr}^2+(\omega-\Delta)^2}\right].
\end{split}
\end{equation}
Then we use Eq. (\ref{important_integral}) and find
\begin{equation}
\begin{split}
\frac{d\tilde{\sigma}_{xy}^{(1)}}{d\Delta}=-& \frac{e^2{\rm
sgn}(eB)}{2\pi^3T}\,{\rm Re}\left[
\Psi^\prime\left(\frac{\gamma_{\rm tr}+i(\mu+\Delta)}{2\pi
T}+\frac{1}{2}\right) \right.
\\& \left.-\Psi^\prime
\left(\frac{\gamma_{\rm tr}+i(\mu-\Delta)}{2\pi T}
+\frac{1}{2}\right)\right].
\end{split}
\end{equation}
Therefore
\begin{equation} \label{dc_Hall_largeB}
\begin{split}
\tilde{\sigma}_{xy}^{(1)}=-& \frac{e^2{\rm sgn}(eB)}{\pi^2}\,{\rm
Im}\left[ \Psi\left(\frac{\gamma_{\rm tr}+i(\mu+\Delta)}{2\pi
T}+\frac{1}{2}\right) \right. \\& \left.+
\Psi\left(\frac{\gamma_{\rm tr}+i(\mu-\Delta)}{2\pi
T}+\frac{1}{2}\right)\right].
\end{split}
\end{equation}
Here we took into account the fact that because of the condition
$\tilde{\sigma}_{xy}(\Delta=\infty)=0$ [see Eq. (\ref{sigma_xy-B=infty})]
and the known asymptotics of the $\Psi$-function, the integration
constant equals zero.

As to the integrals of the two first terms in square brackets in
Eq. (\ref{sigma_xy-B=infty}), they are calculated by means of the formula
\begin{equation}  \label{second_important_int}
\begin{split}
& \int\limits_{-\infty}^\infty
\frac{d\omega}{\cosh^2\frac{\omega-\mu}{2T}}\frac{\omega-\Delta}{\gamma_{\rm
tr}^2+(\omega-\Delta)^2}
\\ & =-\frac{2}{\pi}\,{\rm Im}\Psi^\prime\left(\frac{\gamma_{\rm tr}+i(\mu-\Delta)}{2\pi T}
+\frac{1}{2}\right).
\end{split}
\end{equation}
Its derivation is as follows:
\begin{equation}
\begin{split}
&\int\limits_{-\infty}^\infty
\frac{d\omega}{\cosh^2\frac{\omega-\mu}{2T}}\frac{\omega-\Delta}{\gamma_{\rm tr}^2+(\omega-\Delta)^2}\\
&={\rm Re}\int\limits_{-\infty}^\infty
\frac{d\omega}{\cosh^2\frac{\omega-\mu}{2T}}\frac{1}{\omega-\Delta+i\gamma_{\rm tr}}\\
& ={\rm Im}\int\limits_0^\infty dt\,e^{-\gamma_{\rm tr}
t}\int\limits_{-\infty}^\infty \frac{d\omega\,
e^{it(\omega-\Delta)}}{\cosh^2\frac{\omega-\mu}{2T}}\\
&=2T{\rm Im}\int\limits_0^\infty dt\,e^{- t[\gamma_{\rm
tr}-i(\mu-\Delta)]}\int
\limits_{-\infty}^\infty\frac{dx}{\cosh^2x}\,e^{i2Ttx}\\
&=4T{\rm
Im}\int\limits_0^\infty dt\,e^{-
t[\gamma_{\rm tr}-i(\mu-\Delta)]}\int\limits_{0}^\infty\frac{dx\cos(2Ttx)}{\cosh^2x}\\
&=4\pi T^2\int\limits_0^\infty \frac{dt t\sin[(\mu-\Delta)t]\,e^{-
t\gamma_{\rm tr}}} {\sinh(\pi Tt)}\\
&=-4\pi T^2\frac{d}{d\gamma_{\rm tr}}\int\limits_0^\infty
\frac{dt\sin[(\mu-\Delta)t]\,e^{- t\gamma_{\rm tr}}} {\sinh(\pi
Tt)}\\
&=-\frac{2}{\pi}{\rm Im}\Psi^\prime\left(\frac{\gamma_{\rm
tr}+i(\mu-\Delta)}{2\pi T} +\frac{1}{2}\right),
\label{second_important_int1}
\end{split}
\end{equation}
where, in the very last equality, Eq. (4.131.3) in
Ref.~\cite{Gradstein_Ryzhik} was used. Combining this contribution with that in
Eq. (\ref{dc_Hall_largeB}), we arrive at Eq.~(\ref{B.Hallcond-Delta}).

\section{$U(4)$ symmetry}
\label{C}

The $U(4)$ symmetry in graphene is discussed for example in Appendix A in
Ref. \cite{Gorbar2002PRB}. Here we will describe the properties of this
symmetry used in the main body of the paper.

The 16 generators of the $U(4)$ are
\begin{equation}
\frac{\sigma^\alpha}{2}\otimes I_4,\quad
\frac{\sigma^\alpha}{2i}\otimes\gamma^3,\quad
\frac{\sigma^\alpha}{2}\otimes\gamma^5,\quad\mbox{and}\quad
\frac{\sigma^\alpha}{2}\otimes\frac{1}{2}[\gamma^3,\gamma^5],
\end{equation}
where $I_4$ is the $4 \times 4$ Dirac unit matrix and
$\sigma^\alpha$, with $\alpha=0, 1, 2, 3$, are
four Pauli matrices
connected with spin degrees of freedom [$\sigma^0$ is the $2 \times 2$ unit matrix].
The Dirac matrices are connected with degrees of freedom reflecting
the band structure of graphene with two inequivalent
Dirac points at the corners of the Brillouin zone. In the representation
used in the present paper (see Sec. \ref{2}), the Dirac matrices $\gamma^3$
and $\gamma^5$ are:
\begin{equation}
\gamma^3= i\left(\begin{array}{cc} 0& I \\ I& 0\end{array}\right),\quad
\gamma^5= i\left(\begin{array}{cc} 0&I\\-I&0\end{array}\right),
\label{35}
\end{equation}
where $I$ is the $2\times 2$ unit matrix. The order parameter connected
with the generation of the excitonic gap $\Delta$ is
$\sigma = -\VEV{\overline{\Psi}\Psi} \equiv -\VEV{\Psi^{\dagger}\gamma^0 \Psi}$
\cite{Gusynin1995PRD,Khveshchenko2001PRL,Gorbar2002PRB},
where the Dirac matrix $\gamma^0$ anticommutes both with $\gamma^3$ and
$\gamma^5$ and commutes with $[\gamma^3,\gamma^5]$. The nonzero expectation value
$\VEV{\overline{\Psi}\Psi}=\VEV{[\psi_{KA\sigma}^\dagger \psi_{KA\sigma} + \psi_{K^\prime
A\sigma}^\dagger \psi_{K^\prime A\sigma} - \psi_{KB\sigma}^\dagger
\psi_{KB\sigma} - \psi_{K^\prime B\sigma}^\dagger \psi_{K^\prime
B\sigma}]}$ is directly related to
the electron density imbalance between $A$ and $B$ sublattices of the bipartite hexagonal
lattice of the graphene sheet \cite{Khveshchenko2001PRL,Khveshchenko2004nb}.
The dynamical generation of the gap leads to the spontaneous breakdown of
the $U(4)$ down to the $U(2)_a\times U(2)_b$ with the generators
\begin{equation}
\frac{\sigma^\alpha}{2}\otimes I_4,\quad
\frac{\sigma^\alpha}{2}\otimes \frac{1}{2}[\gamma^3,\gamma^5]
\label{ab}
\end{equation}
[note that, as one can see from Eq. (\ref{35}),
$[\gamma^3,\gamma^5]$ is diagonal]. As to the Zeeman term,
it is connected with the spin density operator $\Psi^{\dagger}\sigma^3\Psi$. Therefore
this term explicitly breaks the $U(4)$ down to the $U(2)_c\times U(2)_d$
with the generators
\begin{equation}
\frac{\sigma^{\alpha^{\prime}}}{2}\otimes I_4,\quad
\frac{\sigma^{\alpha^{\prime}}}{2i}\otimes \gamma^3,\quad
\frac{\sigma^{\alpha^{\prime}}}{2} \otimes \gamma^5,\quad\mbox{and}\quad
\frac{\sigma^{\alpha^{\prime}}}{2}\otimes \frac{1}{2}[\gamma^3,\gamma^5],
\label{cd}
\end{equation}
where $\alpha^{\prime} = 0, 3$. Eqs. (\ref{ab}) and (\ref{cd}) imply that
the Zeeman term and
the generation of the gap together break the $U(4)$ down
to the $U(1)_1\times U(1)_2\times U(1)_3\times U(1)_4$ with the four diagonal
generators
\begin{equation}
\frac{\sigma^{\alpha^{\prime}}}{2}\otimes I_4 \,,\quad
\frac{\sigma^{\alpha^{\prime}}}{2}\otimes \frac{1}{2}[\gamma^3,\gamma^5]\,;\quad
\alpha^{\prime} = 0, 3.
\label{generators}
\end{equation}


\begin{thebibliography}{99}
\bibitem{footnote1} A possibility of a scenario with a strong coupling dynamics
is discussed in Sec. \ref{4}.
\bibitem{footnote2} The crucial feature underlying the magnetic
catalysis phenomenon is the presence of the LLL with energy $E=0$
containing both electron and hole states. This property of graphene
distinguishes it from $2D$ inversion layers.
\bibitem{footnote}
Because of a nonzero scattering rate $\gamma$, the effect of a small gap, $\Delta\alt\gamma$, is
unobservable and, therefore, the location of the transitions between the plateaus are not expected to coincide
exactly with the point where $\Delta$ vanishes. This leads to uncertainties in the relation between
the theoretical model parameter $\mu$ and the experimental quantity $V_g$.
We use the results in Fig.~\ref{phase-diag}
only to find a qualitatively reasonable relation between them.
\bibitem{footnote3} In fact, as was shown above, the gap
disappears for the values of the chemical potential corresponding
to higher LLs in this dynamics
(see Fig. \ref{Tgap}). Therefore the fact that the degeneracy of
higher LLs is removed only by the Zeeman term in this case is evident.
\end{thebibliography}
\end{document}